\documentclass[aps,epsf,preprint]{revtex4}

\def\1{{\bf 1}}
\def\[{\left[}
\def\]{\right]}
\def\be{\begin{eqnarray}}
\def\ee{\end{eqnarray}}
\def\bm{\begin{matrix}}
\def\em{\end{matrix}}
\def\nn{\nonumber}
\def\({\left(}
\def\){\right)}

\def\eq#1{(\ref{#1})}

\def\o{\omega}

\def\G{{\cal G}}

\def\x{\times}

\def\labels#1{\label{#1}}
\def\edc{\end{document}}

\def\bn{\begin{enumerate}}

\def\en{\end{enumerate}}

\def\ol{\overline}
\def\rd{\sqrt{2}}

\def\ba{\begin{array}}
\def\ea{\end{array}}
\def\bc{\begin{center}}
\def\ec{\end{center}}

\def\ol{\overline}
\def\error#1#2{\pm\tiny\hskip-.2cm\begin{array}{l}#1\\#2\end{array}} 
\def\ul{\underline}
\def\edoc{\end{document}}

\begin{document}

\title{Finite Symmetry of Leptonic Mass Matrices}
\author{C.S. Lam}
\address{Department of Physics, McGill University\\
 Montreal, Q.C., Canada H3A 2T8\\
and\\
Department of Physics and Astronomy, University of British Columbia,  Vancouver, BC, Canada V6T 1Z1 \\
Email: Lam@physics.mcgill.ca}

\begin{abstract}
We search for possible symmetries present in the leptonic mixing data from $SU(3)$ subgroups of order up to 511. Theoretical results based on symmetry are compared with  global fits of experimental data in a chi-squared analysis, yielding the following results. There is no longer a  group that can produce all the mixing data without a free parameter, but  a number of them  can accommodate the first or the second column of the mixing matrix.
The only group that fits the third column is $\Delta(150)$. It predicts $\sin^22\theta_{13}=0.11$ and $\sin^22\theta_{23}=0.94$, in good agreement with experimental results.
\end{abstract}
\narrowtext
\maketitle

\section{Introduction}
In the days when the reactor angle of neutrino mixing was thought to be zero and the atmospheric angle maximal, mixings could be taken to be tri-bimaximal, and explained by a $S_4$ symmetry without any free parameter, or groups containing it \cite{LAM1}. Now that both global fits and direct measurements show the reactor angle to be non-negligible and the atmospheric angle  possibly non-maximal \cite{GA,GB,M}, many suggestions have been advanced to explain the new data \cite{T, T2}. In this paper we  investigate whether a finite symmetry still exists to accommodate them. 

The group theory of mixing reviewed in Sec.~II  will be used to extract columns of  possible mixing matrices allowed by a finite group $\G$. With the help of the powerful group software GAP \cite{GAP}, we examine all finite subgroups of $SU(3)$ up to order 511 and compare their predictions with global fits of experimental measurements. Since phases are unknown,  only absolute values of the mixing matrix elements are used.
The experimental data and the global analysis used in a chi-squared comparison will be discussed in Sec.~III. The result presented in Sec.~IV can be summarized as follows. No group can simultaneously accommodate all three columns of the mixing matrix, like $S_4$ was able to do for the tri-bimaximal mixing. Many groups can accommodate the first or the second column, but the success is not necessarily trustworthy  because those globally fitted matrix elements depend on the unknown CP phase. For the third column, only the group $\Delta(150)$ works and it predicts   
 $\sin^22\theta_{13}=0.11$ and $\sin^22\theta_{23}=0.94$, in good agreement with experimental data.
In Sec.~V, the residual symmetry and the corresponding effective mass matrices of  the good groups with $\chi^2<3$ are presented.

We close this section with a brief remark comparing the group theoretical method used here and other approaches. Like texture zeros, both are theories of fermion mass matrices within the Standard Model. There is no need to introduce additional Higgs or valons, together with their alignments. Clebsch-Gordan coefficients are not needed. Unlike the texture-zero approach, mixing parameters are determined by symmetry  here  and not by ratios of fermionic masses. The group-theory approach is consistent with dynamical models based on horizontal symmetry if the valon alignments are given by the invariant eigenvectors of the residual symmetries discussed in the next section. Such alignments  can be obtained from group-invariant potentials in the weak-coupling approximation \cite{LAM2}.

\section{Group Theory of Mixing}

Every  mixing produces a $Z_2\x Z_2$ symmetry in the neutrino Majorana mass matrix and a $Z_n$ symmetry in the left-handed charged-lepton mass matrix. The group theory of mixing to be reviewed below 
\cite{LAM1, LAM2, LAM3} is based on the simple assumption that these natural symmetries are the residual symmetries of a horizontal symmetry group.

Let $\ol M_\nu$ be the symmetric mass matrix of active neutrinos, and $\ol M_e:=M_e^\dagger M_e$ be the hermitian mass matrix of left-handed charged leptons. In the basis where $\ol M_e$ is diagonal, the neutrino mixing matrix $U$ renders $U^T\ol M_\nu U$  diagonal. Let $u_1,u_2,u_3$ be the three columns of $U$. Then the unitary matrices defined by
\be
G_1&=&+u_1u_1^\dagger -u_2u_2^\dagger-u_3u_3^\dagger,\nn\\
G_2&=&-u_1u_1^\dagger +u_2u_2^\dagger-u_3u_3^\dagger,\nn\\
G_3&=&-u_1u_1^\dagger -u_2u_2^\dagger+u_3u_3^\dagger \labels{Gi}\ee
mutually commute and commute with $U$. They satisfy $G_i^2=1$, $G_iG_j=G_k$ ($i,j,k$ different),
and $G_i^T\ol M_\nu G_i=\ol M_\nu$. Thus they generate a $Z_2\x Z_2$ symmetry of $\ol M_\nu$. In the mean time,
since $\ol M_e$ is diagonal, every unitary diagonal matrix $F$ commutes with it, giving $F^\dagger\ol M_eF=\ol M_e$.  If $F^n=1$, then $F$ generates a symmetry group $Z_n$ of $\ol M_e$. We will  assume the eigenvalues of $F$ to be non-degenerate so that $F$ diagonal  forces  $\ol M_e$ to be diagonal. 

Since these symmetries are always present, additional input is required to nail down the mixing. To this end we shall assume $F$ and $G$ to be  residual symmetries of some finite group $\G$, in the sense that both are members of the group. In that case the group structure imposes a correlation between $F$ and $G$, allowing $G$ and hence $U$ to be determined when $F$ is given in its diagonal form.
We call $\G$ a {\it partial-symmetry} group if it contains $F$ and one $G_i$, and a {\it full-symmetry} group if it contains $F$ and two mutually commuting $G_i$'s \cite{T2}. It does not matter which two to choose 
because the third one, being the product of the first two, must also be in $\G$.

Conversely, given a finite group $\G$, any of its order-2 elements is a candidate of $G_i$, and any element with order larger than 2 is a candidate of $F$. In a 3-dimensional irreducible 
{\it unitary} representation of $\G$, and in the basis where $F$ is diagonal, one of the three eigenvectors of $G_i$
is uniquely determined up to normalization. This is the eigenvector with eigenvalue $\pm 1$ if $\det(G_i)=\pm 1$.
This unique eigenvector, which we shall refer to as a {\it mixing vector}, gives rise to one column $u_i$ of the mixing matrix $U$. Note that for this to work, $F$ has to have three distinct eigenvalues. Otherwise $G_i$ is not unique
when $F$ is diagonal, nor is $\ol M_e$ necessarily diagonal when $F$ is.
Going through all combinations of $F$, $G_i$, and 3-dimensional irreducible representations, we get all possible mixing-matrix columns allowed by this group.
As long as $\G$
is a finite group, the number of allowed mixing vectors is finite, though this number could be large for a large group. Two such vectors $u_i$ and $u_j$ may fit into the same mixing matrix $U$ if and only if $G_i$ and $G_j$ commute. 

Within this scheme for Majorana neutrinos, the order of $\G$ must be even because it must contain at least one order-2 member $G_i$. If it is to be a full-symmetry group, then its order must be divisible by 4 because there must be two distinct order-2 elements $G_i$ present. This last condition is necessary but not sufficient  for a full-symmetry group   also requires those three order-2 members to mutually commute.  Please also note that for Dirac neutrinos, there is no need for the order of $\G$ to be even.

Since phases of neutrino mixing have not been measured, only the absolute values of these columns are needed for experimental comparison. Moreover, group theory can never know how to label the neutrino flavor states nor the mass eigenstates, hence any mixing vector from group theory can be used to compare with any of the three columns of the experimental mixing matrix, and the rows may be permuted any way we want before the comparison.

There is a large number of allowed mixing vectors for a large group $\G$ to make the comparison a daunting task to do by hand. 
Fortunately, a powerful free software GAP \cite{GAP} is available to help us. Note however that the irreducible representations given in GAP may not be unitary so to use it we must first obtain from it the corresponding unitary representations.

In Sec~IV, we will use the recipe outlined here to obtain the allowed mixing vectors of all finite subgroups of $SU(3)$ with an order less than 512. 

\section{Experimental Data and Global Fits}

Recent experiments give the following values for the reactor angle $\theta_{13}$ \cite{M}:
\be
\sin^22\theta_{13}&=&0.089\pm0.010\pm0.005\quad{\rm (Daya\ Bay)}\nn\\
\sin^22\theta_{13}&=&0.109\pm0.030\pm0.025\quad{\rm (Double\ Chooz)}\nn\\
\sin^22\theta_{13}&=&0.113\pm0.013\pm0.019\quad{\rm (RENO)}\nn\\
\sin^22\theta_{13}&=&0.104\pm0.060\pm0.045\quad{\rm(T2K,\ normal\ hierarchy)}\nn\\ 
\sin^22\theta_{13}&=&0.128\pm0.070\pm0.055\quad{\rm(T2K,\ inverted\ hierarchy)}.\labels{reactor}\ee
A preliminary result from MINOS \cite{M} also shows that the atmospheric mixing may not be maximal:
\be
\sin^22\theta_{23}&=&0.96\pm 0.04\hskip1.3cm{\rm (MINOS,}\ \nu)\nn\\
\sin^22\theta_{23}&=&0.97\pm 0.03/0.08\quad{\rm (MINOS,}\ \overline{\nu}).\labels{atm}\ee
These results are to some extent anticipated by global fits of the data \cite{GA, GB}. The absolute values of their mixing-matrix elements are shown below, where the errors indicated are 1$\sigma$ errors:

\be
|U^{aN}|&=&\pmatrix{
.814\pm.010 &.558\pm.014&.161\pm.011\cr
.327\error{.160}{.036}&.645\error{.113}{.035}&.691\pm.046\cr
.480\error{.160}{.026}&.522\error{.118}{.044}&.705\pm.045\cr}
\nn\\  \\
|U^{aI}|&=&\pmatrix{
.813\pm.010&.558\pm.014&.164\pm.011\cr
.485\pm.022&.500\pm.041&.718\pm.041\cr
.322\pm.032&.663\pm.031&.676\pm.043\cr
}\nn\\  \\
|U^{bN}|&=&\pmatrix{
.822\pm.010&.547\pm.015&.157\pm.010\cr
.354\error{.098}{.019}&.698\error{.060}{.015}&.623\pm.022\cr
.446\error{.099}{.015}&.462\error{.080}{.022}&.766\pm.018\cr
}\nn\\  \\
|U^{bI}|&=&\pmatrix{
.822\pm.010&.547\pm.015&.157\pm.010\cr
.348\error{.096}{.020}&.694\error{.058}{.017}&.631\pm.025\cr
.451\error{.093}{.016}&.469\error{.078}{.025}&.760\pm.021\cr
},\labels{global}
\ee
where $N$ and $I$ stand for normal and inverted hierarchies, respectively, and $a, b$ are respectively the results taken from \cite{GA} and \cite{GB}. Since phases are not yet measured, only absolute values of the matrix elements are listed and compared. Note that
the 22, 23, 32, 33 matrix elements depend on the unknown CP phase $\delta$, resulting in relatively large errors and may therefore be somewhat unreliable. 

We use the chi-square measure
\be 
\chi^2=\sum_{i=1}^3 (|c_i|-|U_{ij}|)^2/2{\sigma^\pm_{ij}}^2 \labels{chi2}\ee
to gauge the goodness of a theoretically predicted mixing vector $c=(c_1,c_2,c_3)^T$, where $|U_{ij}|\pm\sigma^\pm_{ij}$ is taken from
one of the four globally fitted mixing matrices in \eq{global}. A factor of 2 is included in the definition to simulate the two degrees of freedom in a normalized column, but
since it is the relative size of $\chi^2$ that will be invoked, it does not matter whether we drop that factor or not.
The result of these fits for finite subgroups of $SU(3)$ will be discussed in the next section. GAP is used to produce these results, but as remarked in the last section, the irreducible representations given by GAP are not necessarily unitary, so they have to be rendered unitary first before the mixing vectors $c$ can be computed.

To have a standard for comparison, we list in Table I the chi-square of each column of the tri-bimaximal matrix. The absolute values of its third (bimaximal) column is $(0, .707, .707)^T\sim(0,1,1)^T$, that of its second (trimaximal) columns is $(.577, .577, .577)^T\sim(1,1,1)^T$,
 and that of its first column is $(.816, .408, .408)^T\sim(2,1,1)^T$. 

$$\ba{|c|c|cccc|} \hline
{\rm column}&{\rm mixing\ vector}&{\rm aN}&{\rm aI}&{\rm bN}&{\rm bI}\\ \hline
1&(2,1,1)^T&\ul{4.12}&{9.82}&\ul{3.65}&\ul{4.00}\\
2&(1,1,1)^T&\ul{2.85}&\ul{6.62}&{34.7}&{26.1}\\
3&(0,1,1)^T&{110}&{119}&{126}&{122}\\ \hline
\ea$$

\bc Table I. \quad The $\chi^2$-values of the columns of a tri-bimaximal matrix\ec

With the newly measured reactor angle, the third column having a $\chi^2$-value over 100 is clearly unacceptable. The fit to the  first and second columns are much more tolerable, 
but to some extent that may be due to the large errors associated with the unmeasured CP phase appearing in these two columns. 
In what follows we will reject all fits with $\chi^2>7$; those  that survive in Table I  are underlined for easy comparison.
This criterion of $\chi^2>7$ is rather arbitrary, used here as an illutration; a value other than 7 can be used in the same way. If $\chi^2$ in (8) is defined without the factor 2 in the denominator, then we simply have to change the value 7 to 14.

\section{Finite Subgroups of $SU(3)$}
Finite subgroups of $SU(3)$ with a  three-dimensional irreducible representation and an order less than 512 are tabulated in \cite{P}, and reproduced here in Columns A and B of Table II.  Column A gives the designation of a group in the Small Group Library of GAP;  the first of the pair is the order of the group, and the second is the GAP-assigned number among groups of 
that order. Column B gives the popular name of the group, if there is one. If the group is known under different names, then several of these may be given. Column C indicates whether the group contains $A_4$ or $S_4$ as a subgroup. A symbol $\circ$ indicates that it contains $A_4$, and a symbol $\bullet$
indicates that it contains $S_4$, which then must also contain $A_4$.

If $A_4$ is a subgroup, the group must contain the (unnormalized) trimaximal mixing vector $(1,1,1)^T$. If $S_4$ is a subgroup, then it must contain both $(1,1,1)^T$, and $(2,1,1)^T$, with the corresponding $\chi^2$ given in Table I. For $\chi^2<7$, it also constains $(1, \rd, 1)^T$ with a $\chi^2$ given in Table II below. Since these $\chi^2$ may be considered as reasonable, we can use the group as a partial-symmetry group to build a neutrino mass matrix, with $(2,1,1)^T$ in the first column of its mixing matrix, or  $(1,1,1)^T$, or $(1,\rd,1)^T$, in the second column. 
This strategy has been used, for example, in refs.~\cite{T2,LAM2,LAM3}. 
However, we must not use two of them simultaneously, for if we do so then the group becomes a full-symmetry group, and we know by this survey that no full-symmetry group fits the data.

 Two other symbols also appear in Column C.
The symbol $\x$ is used to indicate groups of odd order, which contains no element of order 2, and therefore will be ignored  from now on. The symbol $p$ is used to identify
groups that can only be partial-symmetry groups whatever the data are. These groups do not contain two mutually commuting order-2 elements so they can never serve as a full-symmetry group.

For each group of even order, we compute all its mixing vectors $c=(c_1,c_2,c_3)^T$ using the recipe discussed in Sec.~2, then its $\chi^2$. We reject cases where $\chi^2>7$ for all four
global fits. Otherwise the values $|c_1|, |c_2|, |c_3|$ are listed in Column D, with the minimal $\chi^2$ among the four global fits appearing in Column E, and the corresponding global fit in Column F. 
The column that it fits, namely, $j$ of $|U_{ij}|$ in \eq{chi2}, appears in column G. For groups containing $A_4$ as a proper subgroup, the $(1,1,1)^T$ mixing vector is understood and will not be listed. For groups containing $S_4$ as a proper subgroup, neither the $(2,1,1)^T, (1,\rd,1)^T$, nor the $(1,1,1)^T$  appears explicitly.  The symbol $-$ is used to indicate that there is no fit whatsoever  with $\chi^2<7$.

When the bar is set higher at $\chi^2=3$, the rejected solutions are indicated by the symbol $\x$ in column H.

The results appearing in Tables IIa, IIb, IIc can be summarized as follows. First,
all the mixing vectors shown can be obtained from a $G_i$ with $\det(G_i)=+1$. Though there are many solutions with $\det(G_i)=-1$, there are no new ones other than those listd in these Tables.
Secondly,
with so many groups and so many possible mixing vectors for each group, it is somewhat surprising that so few passes the experimental test. Besides the $(2, 1, 1)^T$ mixing of the first column, and the $(1,\rd,1)$ mixing of the second column, for groups containing $S_4$, and the $(1, 1, 1)^T$ mixing of the second column for groups containing $A_4$, there are only a few that fit the first or the second column, albeit possibly with a better $\chi^2$. If we set the bar higher at $\chi^2=3$, then there are even fewer
solutions. For example, only the trimaximal
solution $(1,1,1)^T$ survives for  $S_4$ and $A_4$. 
The only group that really fits the third column is $\Delta(150)$, with a mixing vector $(.170, .607, .777)^T$, which gives rise to $\sin^22\theta_{13}=0.11$ and $\sin^22\theta_{23}=0.94$, in good agreement with direct measurements \cite{M}. The only other group that fits the third column with a $\chi^2<7$ is the group $\Delta(294)$,
but it yields too small a reactor angle with $\sin^22\theta_{13}=0.06$ and $\sin^22\theta_{23}=0.97$, with a $\chi^2$ larger than 3.

$$\begin{array}{|c |c|c|c|c|c|c|c|}
\hline
A&B&C&D&E&F&G&H\\ \hline
\[12,3\]&A_4,T&\circ&[.577,.577,.577]&2.85&aN&2 &\\
\[21,1\]&T_7&\x&&&& &\\
\[24,12\]&S_4,O,\Delta(24)&\bullet&[.816,.408,.408]&3.65&bN&1 &\x\\
&&&[.500,.707,.500]&4.95&bI&2 &\x\\
\[27,3\]&\Delta(27)&\x&&&& &\\
\[39,1\]&T_{13}&\x&&&& &\\
\[48,3\]&\Delta(48)&\circ&&&& &\\
\[54,8\]&\Delta(54)&p&[.500,.707,.500]&4.95&bI&2 &\x\\
\[57,1\]&T_{19}&\x&&&& &\\
\[60,5\]&A_5,I,\Sigma(60)&\circ&[.526, .602, .602]&3.68&aN&2 &\x\\
\[75,2\]&\Delta(75)&\x&&&& &\\
\[81,9\]&&\x&&&& &\\
\[84,11\]&&\circ&&&& &\\
\[93,1\]&T_{31}&\x&&&& &\\
\[96,64\]&\Delta(96)&\bullet&&&& &\\
\[108,15\]&\Sigma(36\varphi)&p&-&-&-&- &\\
\[108,22\]&\Delta(108)&\circ&&&& &\\
\[111,1\]&T_{37}&\x&&&& &\\
\[129,1\]&T_{43}&\x&&&& &\\
\[147,1\]&T_{49}&\x&&&& &\\
\[147,5\]&\Delta(147)&\x&&&& &\\
\hline\end{array}$$

\bc Table IIa.\quad Comparison of predictions of $SU(3)$ subgroups with experimental data\ec
\newpage
$$\begin{array}{|c |c|c|c|c|c|c|c|}
\hline
A&B&C&D&E&F&G&H\\ \hline
\[150,5\]&\Delta(150)&p&[.812,.332,.480]&.018&aN&1& \\
&&&[.812,.480,.332]&.086&aI&1 &\\
&&&[.500, .707, .500]&4.95&bI&2 &\x\\
&&&[.170,.607,.777]&1.25&bN&3 &\\
\[156,14\]&&\circ&&&& &\\
\[162,14\]&&p&[.804,.279,.525]&1.41&aN&1 &\\
&&&[.804,.525,.279]&3.05&aI&1 &\x\\
&&&[.500,.707,.500]&4.95&bI&2 &\x\\
\[168,42\]&\Sigma(168), PSL(3,2)&\bullet&[.815,.363,.452]&.267&bN&1 &\\
&&&[.815,.452,.363]&.269&bI&1 &\\
\[183,1\]&T_{61}&\x&&&& &\\
\[189,8\]&&\x&&&& &\\
\[192,3\]&\Delta(192)&\circ&&&& &\\
\[201,1\]&T_{67}&\x&&&& &\\
\[216,88\]&\Sigma(72\varphi)&p&-&-&-&- &\\
\[216,95\]&\Delta(216)&\bullet&&&& &\\
\[219,1\]&T_{73}&\x&&&& &\\
\[228,11\]&&\circ&&&& &\\
\[237,1\]&T_{79}&\x&&&& &\\
\[243,26\]&\Delta(243)&\x&&&& &\\
\[273,3\]&T_{91}&\x&&&& &\\
\[273,4\]&T'_{91}&\x&&&& &\\
\[291,1\]&T_{97}&\x&&&& &\\
\[294,7\]&\Delta(294)&p&[.814,.460,.354]&1.16&aI&1 &\\
&&&[.814,.354,.460]&.312&bI&1 &\\
&&&[.796, .241, .555]&4.63&aN&1 &\x\\
&&&[.500,.707, .500]&4.95&bI&2 &\x\\
&&&[.122,.638,.760]&5.80&bI&3 &\x\\
\hline\end{array}$$

\bc Table IIb.\quad Comparison of predictions of $SU(3)$ subgroups with experimental data\ec
\newpage
$$\begin{array}{|c |c|c|c|c|c|c|c|}
\hline
A&B&C&D&E&F&G&H\\ \hline
\[300,43\]&\Delta(300)&\circ&&&& &\\
\[309,1\]&T_{103}&\x&&&& &\\
\[324,50\]&&\circ&&&& &\\
\[327,1\]&T_{109}&\x&&&& &\\
\[336,57\]&&\circ&&&& &\\
\[351,8\]&&\x&&&& &\\
\[363,2\]&\Delta(363)&\x&&&& &\\
\[372,11\]&&\circ&&&& &\\
\[381,1\]&T_{127}&\x&&&& &\\
\[384,568\]&\Delta(384)&\bullet&[.810,.312,.497]&.188&aN&1 &\\
&&&[.810,.497,.312]&.287&aI&1 &\\
\[399,3\]&T_{133}&\x&&&& &\\
\[399,4\]&T'_{193}&\x&&&& &\\
\[417,1\]&T_{139}&\x&&&& &\\
\[432,103\]&\Delta(432)&\circ&&&& &\\
\[444,14\]&&\circ&&&& &\\
\[453,1\]&T_{151}&\x&&&& &\\
\[471,1\]&&\x&&&& &\\
\[486,61\]&\Delta(486)&p&[.804,.279,.525]&1.41&aN&1 &\\
&&&[.804, .525, .279]&3.05&aI&1 &\x\\
&&&[.500, .707, .500]&4.95&bI&2 &\x\\
\[489,1\]&T_{163}&\x&&&& &\\
\[507,1\]&T_{169}&\x&&&& &\\
\[507,5\]&\Delta(507)&\x&&&& &\\
\hline\end{array}$$

\bc Table IIc.\quad Comparison of predictions of $SU(3)$ subgroups with experimental data\ec

\section{Residual Symmetry and Mass Matrix}
The left-handed mass matrices $\ol M_e$ and $\ol M_\nu$ are determined by the residual symmetry $F$ and $G$, together with
the invariant conditions $F^\dagger \ol M_eF=\ol M_e$ and $G^T\ol M_\nu G=\ol M_\nu$, { provided}
$F$ and $G$ are unitary. 
 These matrices are given in the present section  for the good fits
of Tables IIa, IIb, IIc with $\chi^2<3$. The mixing
vectors in column D of these tables are denoted by $v$, and $\o_n:=\exp(2\pi i/n)$.

Because of phase ambiguity, there are often several $(F,G)$ pairs that yield the same $v$. 
When that happens  only one such pair is given below. Moreover, even that $(F,G)$ is not unique because
a similarity transformation can be applied to the pair to alter both of them, though none of these will alter the mixing vector $v$.

In most cases these matrices are expressed in the representation given in GAP, rather than
the  more familiar $F$-diagonal representation because the latter is usually very complicated. These
GAP representations may or may not be unitary, if not,  which is the case for $\Sigma(168)$,
we must first obtain the unitary form of $F$ and $G$ before proceeding to use the invariant conditions to
calculate the mass matrices.

In the case of $A_4$, the $F$-diagonal representations  are also given because they are more familiar.
For $\Sigma(168)$, the unitary $F$ and $G$ are given in the $F$-diagonal form for reasons
that will be explained later.

Since $\ol M_e$ is hermitian and $\ol M_\nu$ is symmetric, they can be parametrized as
{\small\be
\ol M_e=\pmatrix{\alpha&\beta&\gamma\cr \beta^*&\delta&\epsilon\cr \gamma^*&\epsilon^*&\phi\cr},
\qquad  \ol M_\nu=\pmatrix{a&b&c\cr b&d&e\cr c&e&f\cr},\labels{mm}\ee}
where $\alpha, \delta, \phi$ are real and the rest of the parameters are generally complex. 

A common $F$ that occurs frequently is $F_1$,  of order 3:
{\small\be F_1=\pmatrix{ 0&0&1\cr 1&0&0\cr 0&1&0\cr}.\ 
\labels{F12}\ee}
The corresponding $\ol M_e$ is
{\small \be \ol M_{e}'=F_1^\dagger \ol M_{e}'F_1=
\pmatrix{\alpha&\beta&\beta^*\cr \beta^*&\alpha&\beta\cr \beta&\beta^*&\alpha\cr}.\ee}
It contains three real numbers, $\alpha, \Re(\beta), \Im(\beta)$, just enough to fit
the three charged-lepton masses. For other $F$'s, $\ol M_e$ is different, but
it is still parametrized by three real parameters.

In constrast, as we shall see below, $\ol M_\nu$ is parametrized by four independent complex
parameters. One of them fixes the remaining mixing after a $G_i$-symmetry
is imposed, and the remaining three can be used to fit the three neutrino masses and Majorana phases
(one of which is an unmeasurable overall phase).

\subsection{$A_4$ and Groups Containing $A_4$}
\subsubsection{$v=[.577, .577, .577]$}
{\small\be F=F_1,\quad G=\pmatrix{-1& 0& 0 \cr 0& -1& 0 \cr 0& 0& 1 \cr},\quad
\ol M_e=\ol M_e',\quad \ol M_\nu=\pmatrix{a&b&0\cr b&d&0\cr 0&0&f\cr}.\ee}
Alternatively, in the $F$-diagonal representation,
{\small\be F&=&\pmatrix{1&0&0\cr 0&\o_3&0\cr 0&0&\o_3^2\cr},\quad
G={1\over 3}\pmatrix{-1& 2& 2\cr  2& -1& 2\cr 
  2& 2& -1\cr},\nn\\ \nn\\
 \ol M_e&=&\pmatrix{\alpha&0&0\cr 0&\delta&0\cr 0&0&\phi\cr},\quad
\ol M_\nu=\pmatrix{a&b&c\cr b&a+c-e&e\cr c&e&a+b-e\cr}.
\ee}

%
%
%

\subsection{$\Delta(150)$}
\subsubsection{$v=[.812, .332, .480]$ and $[.812, .480, .332]$}
{\small\be F=F_1,\quad G=-\pmatrix{0&\o_5^4& 0 \cr \o_5& 0& 0 \cr 0& 0& 1\cr},\quad
\ol M_e=\ol M_e',\quad \ol M_\nu=\pmatrix{a&b&c\cr b&a\o_5^3&c\o_5^4\cr c&c\o_5^4&f\cr}
\ee}
\subsubsection{$v=[.170, .607, .777]$}
{\small\be F=F_1,\quad G=-\pmatrix{ 0&\o_5^3& 0 \cr \o_5^2& 0& 0 \cr 0& 0&1\cr},\quad
\ol M_e=\ol M_e', \quad \ol M_\nu=\pmatrix{a&b&c\cr b&a\o_5&c\o_5^3\cr c&c\o_5^3&f\cr}
\ee}

\subsection{$[162,14]$}
\subsubsection{$v=[.804, .279, .525]$}
{\small\be F&=&\pmatrix{ 0& \o_9^7&0 \cr 0& 0& \o_9^2 \cr 1& 0& 0},\quad
G=-\pmatrix{1& 0& 0 \cr 0& 0& 1\cr 0,&1& 0\cr},\nn\\ \nn\\
\ol M_e&=&\pmatrix{\alpha&\beta&\o_9^7\beta^*\cr \beta^*&\alpha&\o_9^4\beta\cr \o_9^2\beta&\o_9^5\beta^*&\alpha\cr},\quad
\ol M_\nu=\pmatrix{a&b&b\cr b&d&e\cr b&e&d\cr}.
\ee}
Note that $F^3=1$ in this case.

\subsection{$\Sigma(168)$}
\subsubsection{$v=[.815, .363, .452]$ and $[.815, .452, .363]$}
This mixing vector has previously been obtained in the first reference of \cite{T2}.

Let $x=(\o_7+\o_7^2+\o_7^4)/2$. Then
{\small \be F={1\over 2}\pmatrix{-x-1&-2x&x-1\cr -x+1&-2x-2&x+1\cr x+1&-2x-2&-x+1\cr},\quad G={1\over 2}\pmatrix{x-1&-2x&-x-1\cr 2x&0&-2x\cr x-1&2x+2&-x-1\cr}\labels{FG168}.
\ee}
It can be checked that $G^2=F^7=1$, but these $F,G$ are not unitary so they cannot be used to obtain the mass matrices. Their unitary representations are analytically  very complicated; actually more complicated in the GAP representation than the $F$-diagonal representation
because a square-root matrix is involved. In the $F$-diagonal
representation, their unitary representations are
{\small\be F=\pmatrix{\o_7^3&0&\cr 0&\o_7^5&0\cr 0&0&\o_7^6},\quad
G_{ij}={1\over 7}G'_{ij}\sqrt{h_j/h_i},
\ee}
where
{\small\be
G'_{11}&=&\o+4 \o^2+2 \o^3+2 \o^4+4 \o^5+\o^6,\nn\\
G'_{12}&=&-4 \o-2 \o^3-3 \o^4-3 \o^5-2 \o^6.\nn\\
G'_{13}&=&-4 \o-4 \o^2-6 \o^4-\o^5-6 \o^6,\nn\\
G'_{21}&=&-4 \o-6 \o^2-6 \o^3-4 \o^4-\o^6,\nn\\
G'_{22}&=&4 \o+2 \o^2+\o^3+\o^4+2 \o^5+4 \o^6\nn\\
G'_{23}&=&-3 \o^2-2 \o^3-4 \o^4-2 \o^5-3 \o^6,\nn\\
G'_{31}&=&-3 \o-4 \o^2-3 \o^3-2 \o^5-2 \o^6,\nn\\
G'_{32}&=&-6 \o-4 \o^2-\o^3-4 \o^4-6 \o^5,\nn\\
G'_{33}&=&2 \o+\o^2+4 \o^3+4 \o^4+\o^5+2 \o^6,\nn\\
h_1&=&-2 \o-(4/3) \o^2-(4/3) \o^3-(4/3) \o^4-(4/3) \o^5-2 \o^6,\nn\\
h_2&=&-(4/3) \o-(4/3) \o^2-2 \o^3-2 \o^4-(4/3) \o^5-(4/3) \o^6,\nn\\
h_3&=&-(4/3) \o-2 \o^2-(4/3) \o^3-(4/3) \o^4-2 \o^5-(4/3) \o^6,
\ee}
and $\o:=\o_7$. Though it may not be obvious, the quantities $h_i$ are positive
and $G$ is unitary, as can be easily verified numerically. The matrix $G'/7$
is the matrix $G$ of \eq{FG168} in the $F$-diagonal representation, and the factors $\sqrt{h_i}$
come from the similarity transformation which renders the representation unitary. 

As to the mass matrices, $\ol M_e$ has to be diagonal in the $F$-diagonal representation, with
$\beta=\gamma=\epsilon=0$ in \eq{mm}. The analytical form of the resulting $\ol M_\nu$
is far too complicated, but the numerical relation of the parameters in \eq{mm} can be obtained from the numerical form 
of $G$ to be
\be 
e &=&(-1.401+1.757i)a+ (.623+2.732i)b+(1.123+1.409i)c-(2.024+.975i)d, \nn\\
f &=&(3.648-1.757i)a+ (1-4.381)b-(1+4.381i)c+(3.148+3.947i)d.
\ee

In the first reference of \cite{T2}, an equivalent but simpler result is obtained in terms of unitary representations. 
Since GAP is used throughout this paper, and since GAP employs only rational and cyclotomic numbers, unitary representation is not always possible. This forces the user to convert the non-unitary GAP representations into unitary
representations, which results in the complicated expressions exhibited above.

\subsection{$\Delta(294)$}

\subsubsection{$v=[.814,.460,.354]$  {\rm and}  $[.814,.354,.460]$}
{\small \be F=F_1,\quad 
G=-\pmatrix{ 0& \o_7^6& 0\cr \o_7& 0& 0\cr  0& 0& 1\cr},\quad \ol M_e=\ol M_e',\quad
\ol M_{\nu}=\pmatrix{a&b&c\cr b&a\o_7^5&c\o_7^6\cr c&c\o_7^6&f\cr}.\ee}


%


\subsection{$\Delta(384)$}

\subsubsection{$v=[.810, .312, .497]$ {\rm and}  $[.810, .497, .312]$}
This mixing vector has previously been obtained in the first reference of \cite{T2}.

{\small\be F=F_1,\quad G=\pmatrix{ -1& 0& 0 \cr 0& 0& \o_8^3 \cr  0& -\o_8& 0\cr},
\quad \ol M_e=\ol M_e',\quad \ol M_\nu=\pmatrix{a&b&b(1-i)/\rd\cr b&d&e\cr b(1-i)/\rd&e&-id\cr}.
\ee}

\subsection{$\Delta(486)$}
\subsubsection{$v=[.804, .279, .525]$}
The solution is identical to case {\bf C} because [162,14] is a subgroup of $[486,61]=\Delta(486)$.

\section{Summary}
We have used the group theory of mixing to determine whether any of the finite subgroups of $SU(3)$ up to order 511 can be a symmetry group of neutrino mixing. We conclude that none could be a full-symmetry group, but several may serve as a partial-symmetry group for column 1 or column 2 of the mixing matrix. Due to the unknown CP phase which
affects the magnitude of the first two columns, it is not clear which of these groups  is a better one.
The only group where the third column can be accommodated is $\Delta(150)$, which yields
$\sin^22\theta_{13}=0.11$ and $\sin^22\theta_{23}=0.94$, in good agreement with direct experimental measurements and global fits. An attemp to construct a dynamical model of $\Delta(150)$  is underway.

I am grateful to Profs.~A. Hulpke   and D. Pasechnik   for their help in using GAP, and to Prof.~John McKay for discussions of finite group theory.

\edoc